\documentclass[twocolumn]{aastex63}
\usepackage{amsmath}
\usepackage{xcolor}

\begin{document}

\title{MAMMOTH-Subaru III. Ly$\alpha$ Halo Extended to $\sim200$ kpc Identified by Stacking $\sim 3300$ Ly$\alpha$ Emitters at $z=2.2-2.3$}

\author{Haibin Zhang}
\affiliation{Department of Astronomy, Tsinghua University, Beijing 100084, People’s Republic of China; \url{haibinzhang@mail.tsinghua.edu.cn}}

\author{Zheng Cai}
\affiliation{Department of Astronomy, Tsinghua University, Beijing 100084, People’s Republic of China; \url{haibinzhang@mail.tsinghua.edu.cn}}

\author{Yongming Liang}
\affiliation{Institute for Cosmic Ray Research, The University of Tokyo, Kashiwa, Chiba 277-8582, Japan}

\author{Ke Ma}
\affiliation{Department of Astronomy, Tsinghua University, Beijing 100084, People’s Republic of China; \url{haibinzhang@mail.tsinghua.edu.cn}}

\author{Nobunari Kashikawa}
\affiliation{National Astronomical Observatory of Japan, 2-21-1 Osawa, Mitaka, Tokyo 181-8588, Japan}
\affiliation{Department of Astronomy, School of Science, The University of Tokyo, 7-3-1 Hongo, Bunkyo-ku, Tokyo 113-0033, Japan}
\affiliation{Research Center for the Early Universe, The University of Tokyo, 7-3-1 Hongo, Bunkyo-ku, Tokyo 113-0033, Japan}

\author{Mingyu Li}
\affiliation{Department of Astronomy, Tsinghua University, Beijing 100084, People’s Republic of China; \url{haibinzhang@mail.tsinghua.edu.cn}}

\author{Yunjing Wu}
\affiliation{Department of Astronomy, Tsinghua University, Beijing 100084, People’s Republic of China; \url{haibinzhang@mail.tsinghua.edu.cn}}
\affiliation{Steward Observatory, University of Arizona, 933 N Cherry Ave, Tucson, AZ 85721, USA}

\author{Qiong Li}
\affiliation{Jodrell Bank Centre for Astrophysics, University of Manchester, Oxford Road, Manchester, UK}

\author{Sean D. Johnson}
\affiliation{Department of Astronomy, University of Michigan, Ann Arbor, MI 48109, USA}

\author{Masami Ouchi}
\affiliation{Institute for Cosmic Ray Research, The University of Tokyo, Kashiwa, Chiba 277-8582, Japan}
\affiliation{National Astronomical Observatory of Japan, 2-21-1 Osawa, Mitaka, Tokyo 181-8588, Japan}
\affiliation{Kavli Institute for the Physics and Mathematics of the Universe (WPI), University of Tokyo, Kashiwa, Chiba 277-8583, Japan}

\author{Xiaohui Fan}
\affiliation{Steward Observatory, University of Arizona, 933 N Cherry Ave, Tucson, AZ 85721, USA}

\begin{abstract}
In this paper, we present a Ly$\alpha$ halo extended to $\sim200$ kpc 
identified by stacking 
$\sim 3300$ Ly$\alpha$ emitters at $z=2.2-2.3$. 
We carry out imaging observations and data reduction with Subaru/Hyper Suprime-Cam (HSC). 
Our total survey area is $\sim12$ deg$^2$ and imaging depths are $25.5-27.0$ mag. Using the imaging data, 
we select 1,240 and 2,101 LAE candidates at $z=2.2$ and 2.3, respectively. We carry out 
spectroscopic observations of our 
LAE candidates and data reduction with Magellan/IMACS to estimate the contamination rate 
of our LAE candidates. 
We find that the contamination rate of our sample is low (8\%). We stack our LAE candidates 
with a median stacking method 
to identify the Ly$\alpha$ halo at $z=2$. We show that the Ly$\alpha$ halo is extended to 
$\sim200$ kpc at a surface brightness level of $10^{-20}$ erg s$^{-1}$ cm$^{-2}$ arcsec$^{-2}$. 
Comparing to previous studies, our Ly$\alpha$ halo is more extended at radii of $\sim25-100$ kpc, 
which is not likely caused by the contamination in our sample but by different redshifts and fields instead. To investigate how central galaxies 
affect 
surrounding LAHs, 
we divide our LAEs into subsamples based on the Ly$\alpha$ luminosity ($L_{\rm Ly\alpha}$), 
rest-frame Ly$\alpha$ equivalent width (EW$_0$), and UV magnitude (M$_{\rm uv}$). 
We stack the subsamples and find that higher $L_{\rm Ly\alpha}$, lower EW$_0$, 
and brighter M$_{\rm uv}$ cause more extended halos. Our results suggest that more 
massive LAEs generally have more extended Ly$\alpha$ halos.
\end{abstract}


\section{Introduction} \label{sec:intro}

In the last two decades, diffuse and faint Ly$\alpha$ emission known as Ly$\alpha$ halos (LAHs) 
are found ubiquitously around Ly$\alpha$ emitters (LAEs) at $z\sim2-7$ 
(e.g. \citealt{hayashino2004, rauch2008, steidel2011, matsuda2012, feldmeier2013, momose2014, momose2016, wisotzki2016, wisotzki2018, xue2017, leclercq2017, cai2019, kakuma2021, niemeyer2022, kikuchihara2022}). 
Some LAHs are identified by stacking a large number of LAEs to increase the signal-to-noise ratio 
(e.g. \citealt{momose2014}, \citealt{wisotzki2018}, and \citealt{niemeyer2022}), 
while the other LAHs are detected individually by deep integral-field-unit (IFU) observations 
(e.g. \citealt{wisotzki2016} and \citealt{leclercq2017}.) LAHs are believed to trace hydrogen gas in 
the circumgalactic medium (CGM), and the CGM is related to the central galaxy via processes 
such as inflow, outflow, and radiation (e.g. Dekel et al. 2009; Sadoun et al. 2019). Thus, 
LAH is an important population to understand how CGM affects galaxy formation and evolution.

Until recent years, LAHs are only identified at radii smaller than $\sim100$ kpc due to the limited depths (e.g. \citealt{momose2014,xue2017,battaia2019,cai2019}). \citet{kakuma2021} and \citet{kikuchihara2022} greatly update the record and find that LAHs are possibly extended to $\sim1000$ kpc, although the radius bin sizes are not small enough to investigate the transition from central galaxy to CGM (several kpc to several hundred kpc). On the other hand, \citet{niemeyer2022} use relatively small radius bin sizes and identify a LAH extended to 320 kpc at $z=1.9-3.5$, although there are only two bins at radii $\gtrsim100$ kpc ($80-160$ and $160-320$ kpc). 

After identification of LAHs, previous studies have investigated how central 
galaxies affect LAHs. \citet{momose2016} find that galaxies with a lower Ly$\alpha$ 
luminosity ($L_{\rm Ly\alpha}$), lower rest-frame Ly$\alpha$ equivalent width (EW$_0$), 
and brighter UV magnitude (M$_{\rm uv}$) have more extended halos. On the contrary, 
\citet{xue2017} and \citet{niemeyer2022} find that galaxies with higher $L_{\rm Ly\alpha}$ 
show more extended halos. The reason causing 
contradictory results is still not clear. 

In this paper, we present a LAH extended to $\sim200$ kpc identified by 
stacking $\sim 3300$ Ly$\alpha$ emitters at $z=2$. This paper is structured as follows. 
We introduce our imaging observations and data reduction in Section \ref{sec:data}. 
The sample selection is presented in Section \ref{sec:selection}. 
Section \ref{sec:spec} shows our spectroscopic observations, data reduction, 
and contamination rate estimation. Our results and discussion are presented in 
Section \ref{sec:results}. Finally we summarize this paper in Section \ref{sec:summary}.

Throughout this paper, we use AB magnitudes \citep{oke1983} and physical distances. A $\Lambda$CDM cosmology with $\Omega_m=0.3$, $\Omega_\Lambda=0.7$, and $h=0.7$ is adopted.

\section{Imaging Observations and Data Reduction} \label{sec:data}

We carry out narrowband (NB) and broadband (BB) imaging observations 
with Subaru/Hyper Suprime-Cam (HSC; \citealt{miyazaki2018, komiyama2018, kawanomoto2018, furusawa2018}). 
The narrowband filters we use are NB387 ($\lambda_{\mathrm{c}}=3863$ \AA, FWHM=55 \AA) 
and NB400 ($\lambda_{\mathrm{c}}=4003$ \AA, FWHM=92 \AA), and the broadband is $g$ ($\lambda_{\mathrm{c}}=4754$ \AA, FWHM=1395 \AA). 
The central wavelengths of 
NB387 and NB400 filters are chosen to detect redshifted Ly$\alpha$ emission 
at $z=2.2$ and $2.3$, respectively. More details of the observations are described in \citet{liang2021} 
and Liang et al. (in prep.). In brief, we observe four fields (J0210, J0222, J0924, and J1419) with 
NB387 and four other fields (J0240, J0755, J1133, and J1349) with NB400 between January 2018 and 
March 2020. The detailed field selection are described in \citet{cai2016}, \citet{liang2021}, 
Cai et al. (in prep.), and Liang et al. (in prep.). The seeing sizes are $0.7-1.2$ arcsec 
depending on the fields, with smallest seeing in J1419 and largest seeing in J0210. The total 
survey area is $\sim12$ deg$^2$.

The imaging data are reduced with the HSC pipeline dubbed \textit{hscPipe} (\citealt{bosch2018, aihara2018}). 
The NB387 and $g$ images of four NB387 fields are reduced by \citet{liang2021}. We reduce the NB400 and $g$ images of 
the four NB400 fields in the same manner as \citet{liang2021}. In brief, we first carry out bias subtraction, dark subtraction, 
flat-field calibration, and stacking of individual exposures. We then use the imaging data from the Panoramic Survey Telescope 
and Rapid Response System 1 (Pan-STARRS1; \citealt{schlafly2012, tonry2012, magnier2013}) survey to calibrate the astrometry 
and photometry with sky subtraction. The $5\sigma$ detection limits of the final imaging products are $24.3-25.0$, $25.5-25.8$, 
and $26.2-27.0$ mag for NB387, NB400, and $g$, respectively. The above detection limits are measured in a $1''.7$ diameter aperture, 
except for J0210 field in a $2''.5$ diameter aperture. Details of the fields are summarized in Table \ref{tab:fields}.

We carry out source detection and photometry with SExtractor \citep{bertin1996} to build our source catalogs. The source 
catalogs of four NB387 fields are obtained by \citet{liang2021}, and we make the source catalogs of the four NB400 fields 
in the same manner. Firstly, we convolve the NB400 and $g$ images with proper Gaussian kernels to match the point-spread-functions (PSFs) 
of NB and BB filters. Then, we use the NB400 images as the reference images to detect sources in the dual-image mode of SExtractor. 
The detection criterion is 15 adjacent pixels above the $1.2\sigma$ limit. To reduce the effect of different depths across the 
field, we use the sky background root-mean square map as the weight image. The mesh size used for sky background estimation is 
128 pixels. We do not use regions with low signal-to-noise ratios (SNRs) or bad pixels such as field edges and bright star vicinity.

\begin{deluxetable*}{cccccccc}
\tablenum{1}
\tablecaption{Information of fields in this study\label{tab:fields}}
\tablewidth{0pt}
\tablehead{
\colhead{Field} & \colhead{R.A. (J2000)} & \colhead{Decl. (J2000)} & \colhead{Obs. Date} & \colhead{Filters} & \colhead{$m_{\mathrm{NB}, 5 \sigma}$} & \colhead{$m_{\mathrm{BB}, 5 \sigma}$} &\colhead{$N_{\rm LAE}$} \\
\colhead{name} & \colhead{hh:mm:ss} & \colhead{dd:mm:ss} & \colhead{month,year} & \colhead{name} &
\colhead{mag} & \colhead{mag} & count
}
\decimalcolnumbers
\startdata
J0210 & 02:09:58.90 & +00:53:43.0 & Jan., 2018         & NB387 \& $g$ & 24.25     & 26.34  & 227   \\
J0222 & 02:22:24.66 & -02:23:41.2 & Jan., 2018         & NB387 \& $g$ & 24.99     & 27.01  & 422   \\
J0924 & 09:24:00.70 & +15:04:16.7 & Jan. \& Mar., 2019 & NB387 \& $g$ & 24.74     & 26.63  & 311   \\
J1419 & 14:19:33.80 & +05:00:17.2 & Mar., 2019         & NB387 \& $g$ & 24.81     & 26.80  & 280    \\
J0240 & 02:40:05.11 & -05:21:06.7 & Nov., 2019        & NB400 \& $g$ & 25.61     & 26.80   & 517   \\
J0755 & 07:55:35.89 & +31:09:56.9  & Nov., 2019        & NB400 \& $g$ & 25.83     & 26.50  & 545    \\
J1133 & 11:33:02.40 & +10:05:06.0  &  Mar., 2020       & NB400 \& $g$ & 25.49     & 26.30  & 403    \\
J1349 & 13:49:40.80 & +24:28:48.0  &  Mar., 2020        & NB400 \& $g$ & 25.67     & 26.15  & 636  \\
\enddata
\tablecomments{Column 1: field name; Column 2: right ascension; Column 3: declination; Column 4: date of observation; Column 5: filters used; Column 6: 5$\sigma$ limiting magnitude of narrowband; Column 7: 5$\sigma$ limiting magnitude of broadband; Column 8: number of LAE candidates selected in this study. }
\end{deluxetable*}

\section{Sample Selection} \label{sec:selection}

Using the reduced images and source catalogs, we select LAEs by an excess of the narrowband minus 
broadband (NB-BB) color in a manner similar to previous studies (e.g. \citealt{konno2016, liang2021}). 
Our NB387 LAE selection is based on the LAE catalogs in \citet{liang2021}, but we apply more strict 
criteria to mainly reduce spurious faint sources. Our selection criteria for NB387 LAEs are: 

\begin{equation} \label{eq:NB387}
\begin{gathered}
g_{\rm ap}-NB387_{\rm ap}>0.3\ \mathrm{and}\ g_{\rm tot}-NB387_{\rm tot}>0.3\ \mathrm{and}\\
g_{\rm ap}-NB387_{\rm ap}>(g_{\rm ap}-NB387_{\rm ap})^{3\sigma}\ \mathrm{and}\\
g_{\rm tot}-NB387_{\rm tot}>(g_{\rm tot}-NB387_{\rm tot})^{3\sigma}\ \mathrm{and}\\
20.5<NB387_{\rm ap}<NB387^{5\sigma},
\end{gathered}
\end{equation}

where the subscripts ``ap" and ``tot" mean the aperture and total magnitudes, respectively. 
We use the ``AUTO'' magnitude in SExtrator as the total magnitude. The superscripts ``$3\sigma$'' and ``$5\sigma$'' 
represent the $3\sigma$ and $5\sigma$ limits, respectively. If the $g$ magnitude of a source is fainter than 
the $2\sigma$ limit, we use the $2\sigma$ limit instead. The $(g-NB387)^{3\sigma}$ is the $3\sigma$ color error 
that is calculated by $(g-NB387)^{3\sigma}=-2.5\log(1\pm3\sqrt{f_{\rm{err},g}^2+f_{\rm{err},\rm{NB387}}^2}/f_{\rm NB387})$, 
where $f_{\rm{err},g}$ and $f_{\rm{err},\rm{NB387}}$ are the $1\sigma$ errors in $g$ and NB387, respectively. The NB-BB color 
limit ($>0.3$) corresponds to a rest-frame Ly$\alpha$ equivalent width of $>20${\AA}. After applying the above criteria, 
we carry out visual inspection to remove spurious sources such as cosmic rays and satellite trails. After visual inspection, 
the final number of our NB387 LAE candidates is 1240.

\begin{figure}[htb]
\plotone{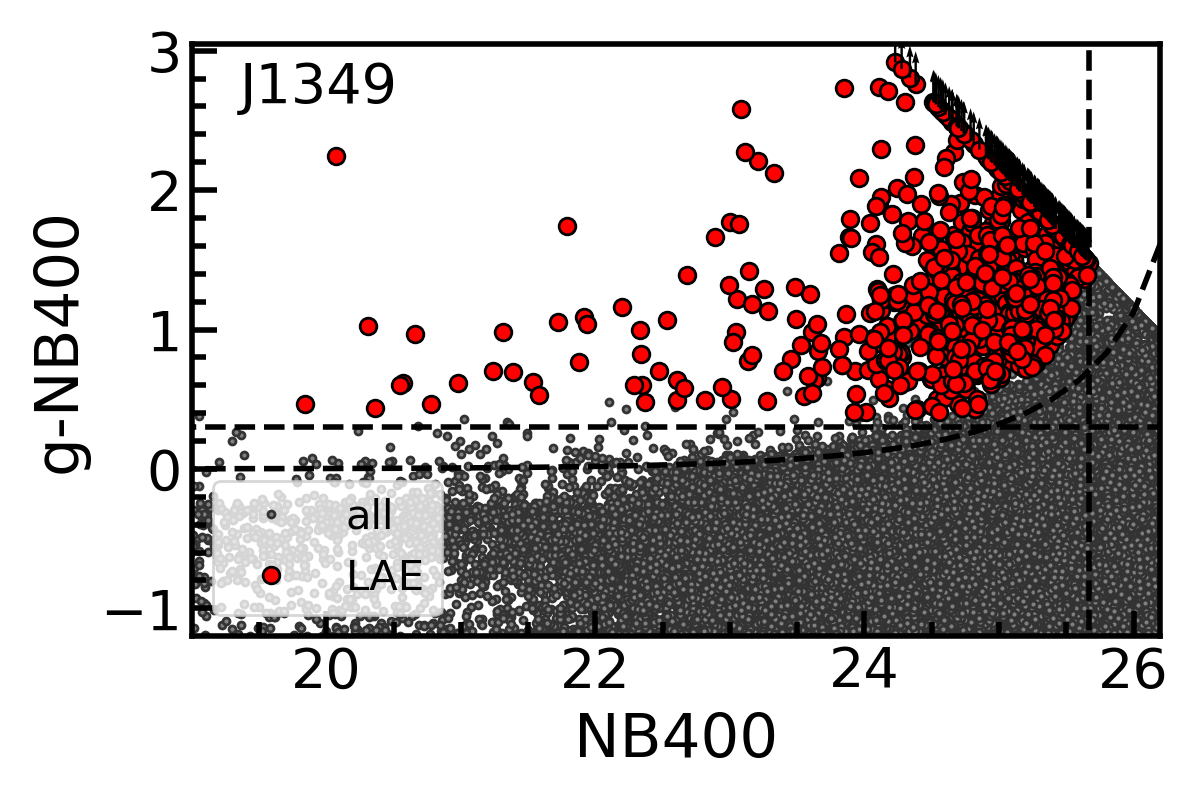}
\caption{Color-magnitude diagram of NB400 LAE candidates (red) and all detected sources in the source catalog (black). 
The all detected sources include contamination such as foreground objects, cosmic rays, and satellite trails. 
The dashed lines represent our selection criteria. This figure shows the LAE selection in the J1349 field as an example. There is 
a sequence of LAE candidates with black arrows in the upper right corner. This is because that if the $g$ magnitude of a source is 
fainter than the $2\sigma$ limit, we use the $2\sigma$ limit instead.}
\label{fig:selection}
\end{figure}

We select NB400 LAEs in the same manner as NB387 LAEs. As shown in Figure \ref{fig:selection},
our selection criteria for NB400 LAEs are:
\begin{equation} \label{eq:NB400}
\begin{gathered}
g_{\rm ap}-NB400_{\rm ap}>0.4\ \mathrm{and}\ g_{\rm tot}-NB400_{\rm tot}>0.4\ \mathrm{and}\\
g_{\rm ap}-NB400_{\rm ap}>(g_{\rm ap}-NB400_{\rm ap})^{3\sigma}\ \mathrm{and}\\
g_{\rm tot}-NB400_{\rm tot}>(g_{\rm tot}-NB400_{\rm tot})^{3\sigma}\ \mathrm{and}\\
18.0<NB400_{\rm ap}<NB387^{5\sigma},
\end{gathered}
\end{equation}
where the meaning of symbols is the same as Equation \ref{eq:NB387}. After visual inspection, 
the final number of our NB400 LAE candidates is 2101. 

\section{Contamination Rate Estimation}\label{sec:spec}

\subsection{Spectroscopic Observations and Data Reduction} 

To estimate the contamination rate of our NB387 and NB400 LAE candidates, we carry out spectroscopic 
observations with Magellan/IMACS on September 29 and 30, 2022. The Magellan/IMACS is set in the multislit 
spectroscopy mode with the f/2 camera, which provides a field of view of $12''$ in circular radius. We use the 
400 lines per mm grism combined with a filter to cover the wavelength between $\sim3600$ and 5700 {\AA}. 
The spectral resolution is $\sim7${\AA} using this setup with a slit width of 1.2 arcsec and slit length 
of 8.0 arcsec. We observe one pointing in each of J0210 and J0222 fields for NB387 LAEs, and 2 pointings in 
the J0240 field for NB400 LAEs. The on-source exposure times are 7500s and 6000s for NB387 and NB400 LAEs, 
respectively. 

We reduce the spectroscopic data using the official pipeline named \textit{COSMOS}. We carry out bias 
subtraction, flat calibration, wavelength calibration, sky subtraction, two-dimensional spectrum extraction, 
and stacking of individual exposures. We use dome flat frames for the flat calibration, and Helium-Mercury 
lamp spectra for the wavelength calibration. We do not conduct flux calibration yet, because it is not 
necessary for the purpose of contamination rate estimation. 

\subsection{Contamination Rate Calculation}
After the data reduction, we obtain 120 and 151 spectra for NB387 and NB400 LAEs, respectively.
Among the total 271 spectra, there are 120 spectra with a detected emission line at the expected 
wavelengths of Ly$\alpha$. The detection fraction is not high because that the real on-source exposure 
time is shorter than expected. Another reason is that we also observe low-priority faint LAE candidates 
as the available slits on a slit mask are much more than our high-priority bright targets. Among the 
120 spectra with detection, 22 spectra are confirmed LAEs at $z=2$ with $\geq2$ emission lines, and 
2 spectra are foreground objects with $\geq2$ emission lines. The 22 LAEs with $\geq2$ emission lines are typically detected in Ly$\alpha$, {\sc C iv}, and/or He {\sc ii}. 
The 2 foreground objects with $\geq2$ emission lines include a {\sc C iv}+{\sc C iii} emitter at $z=1.711$ and a {\sc C iv}+He {\sc ii}+{\sc C iii} emitter at $z=1.793$. 
Because our spectral resolution ($\sim7${\AA}) is not 
high enough to resolve the doublet of a foreground {\sc [O ii]} emitter, 
we only use the spectra with $\geq2$ emission lines when estimating the contamination rate. The 
contamination rate of our LAE candidates is thus 2/(22+2)$\approx$8\%. 
The NB magnitudes of the 24 objects with $\geq2$ emission lines are 21.3-25.3 mag.

\section{Results and Discussion} \label{sec:results}
\subsection{LAE Stacking}\label{sec:stacking}

After obtaining our LAE catalogs, we stack the NB387 and NB400 LAE candidates 
to detect the faint and extended 
Ly$\alpha$ halo at $z=2$. First, we match PSFs of all filters (NB387, NB400, and $g$) in all the 8 fields by 
convolution with proper Gaussian kernels. Then we make cutout narrowband (NB387 or NB400) and broadband ($g$) 
images with a size of $84''\times84''$ for each LAE. During this process, we remove six NB387 LAEs because 
they are too close to the field edges and we cannot make their cutout images with the $84''\times84''$ size. 
We subtract the narrowband images by broadband images to obtain Ly$\alpha$ images. 
We stack the Ly$\alpha$ images with a median stacking method for NB387 and NB400 LAEs separately. 
Because the expected redshifts of NB387 and NB400 LAEs are very close 
($\Delta z\sim0.1$), we also stack NB387 and NB400 LAEs together and this is referred 
to as the \textit{all} sample in the following sections. 
The number of LAEs used for stacking are 1234, 2101, and 3335 for the 
NB387, NB400, and \textit{all} LAEs, respectively. It should be noted that the 3335 \textit{all} LAEs contain 117 Ly$\alpha$ blobs (LABs) whose details are described in Li et al. (in prep.) and Zhang et al. (in prep.). We do not remove these LABs during the stacking, because there are no evidences showing that extended Ly$\alpha$ emission of LABs is generally distinct (\citealt{zhang2020}). After stacking, 
we globally subtract the median value measured in a radius of $34''-42''$ 
($\sim280-350$ kpc) from the image to remove the small sky residual or 
sky over-subtraction. 

We estimate the uncertainty of our stacking results by the following method. 
Firstly, we randomly choose sky regions and make cutout sky images 
with the same number as our LAE samples. The sky region means that a region 
with no detected objects within a radius of 20 pixels ($3''.4$) from the center. 
This is because when we visually inspect LAEs in Section \ref{sec:selection}, 
we only remove LAE candidates contaminated by close (separation $\lesssim 3''$) 
non-LAE objects. As a result, it is natural that far (separation $\gtrsim 3''$) 
contaminants exist around our LAEs. Then we stack the sky images in the same 
manner as our LAEs. We repeat the above procedures for 100 times and obtain 
100 stacked sky images. Finally we plot histograms of surface brightness using the stacked 
sky images and fit a Gaussian function to measure the 1$\sigma$ uncertainties 
at different radii. We find that the 1$\sigma$ uncertainties at radii of $\gtrsim 75$ kpc are between $4\times 10^{-20}$ and $5\times 10^{-20}$ erg s$^{-1}$ cm$^{-2}$ arcsec$^{-2}$. The surface brightness distributions of sky at different radii are shown in Figure \ref{fig:sky_hist} in Appendix. 

\begin{figure}[htb]
\plotone{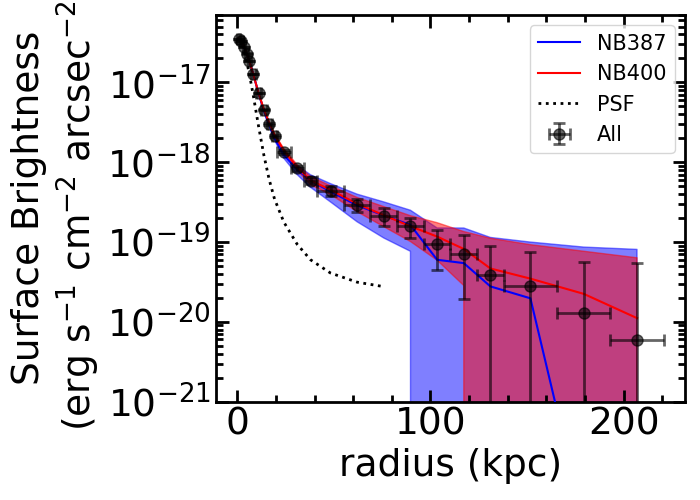}
\caption{Ly$\alpha$ surface brightness profiles of NB387 (blue), NB400 (red), \textit{all} (black) LAEs. The horizontal error bars indicate the ranges of radius bins, while the vertical error bars are the uncertainties of surface brightness obtained by our simulation. The point-spread-function (PSF; dotted line) 
is obtained by stacking 
533 point sources with NB magnitudes of $18-22$ mag. We scale the PSF to the \textit{all} profile at the radius of $\sim0$ kpc for comparison, 
while the NB387, NB400, and \textit{all} profiles are not scaled.}
\label{fig:stacking_1d}
\end{figure}

Figure \ref{fig:stacking_2d} shows our stacked Ly$\alpha$ images and 
Figure \ref{fig:stacking_1d} shows the Ly$\alpha$ surface brightness profiles. 
Clearly, the Ly$\alpha$ emission is extended to $\sim200$ kpc at a surface 
brightness level of $10^{-20}$ erg s$^{-1}$ cm$^{-2}$ arcsec$^{-2}$, 
much more extended than the PSF. The PSF is obtained by stacking 533 point 
sources with NB magnitudes of $18-22$ mag. All of the three profiles 
(NB387, NB400, and \textit{all}) are consistent within the 1$\sigma$ error 
without any scaling.  

\begin{figure}[htb]
\plotone{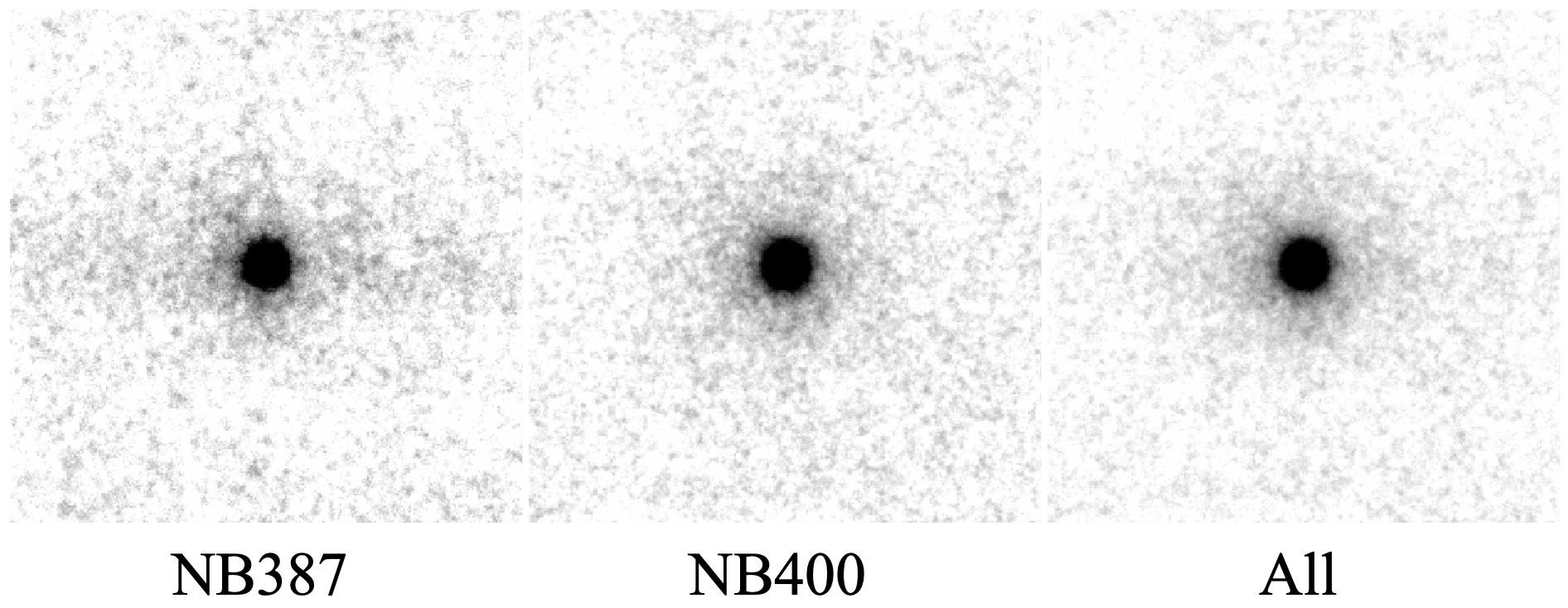}
\caption{Stacked Ly$\alpha$ images of NB387, NB400, \textit{all} LAEs. 
The size of each image is $50''.4\times50''.4$ ($\sim415 \times 415$ kpc$^2$). 
The lower and upper surface brightness limits for plotting the images are 0 
and $2\times10^{-18}$ erg s$^{-1}$ cm$^{-2}$ arcsec$^{-2}$, respectively.}
\label{fig:stacking_2d}
\end{figure}

\begin{figure}[htb]
\plotone{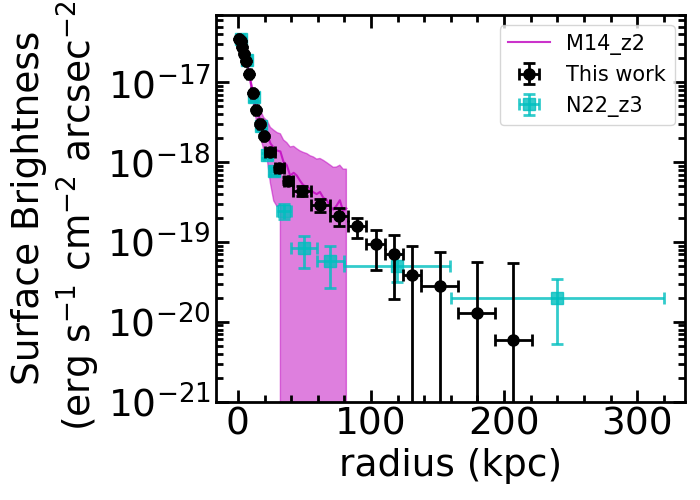}
\caption{Ly$\alpha$ surface brightness profiles of Momose et al. (\citeyear{momose2014}; magenta) and Lujan Niemeyer et al. (\citeyear{niemeyer2022}; cyan) compared with this study (black). We scale the profiles of M14 and N22 to this study at the radius of $\approx0$ kpc 
for comparison.}
\label{fig:halo_compare}
\end{figure}

We compare our stacking result with previous LAH studies at $z=2-3$ as shown 
in Figure \ref{fig:halo_compare}. Using a method similar to this study, 
Momose et al. (\citeyear{momose2014}; hereafter M14) stack 3556 LAEs 
at $z=2.2$ with Subaru/Suprime-Cam and identify LAH extended to $\sim80$ kpc. 
Our result is consistent with M14 after scaling, although the uncertainty of 
M14 is relatively large at radii larger than 30 kpc. We also compare our results with 
Lujan Niemeyer et al. (\citeyear{niemeyer2022}; hereafter N22). 
N22 stack 968 spectroscopically confirmed LAEs at $z=1.9-3.5$ with 
the HETDEX data, and identify Ly$\alpha$ emission extended to 320 kpc. 
Our result is consistent with N22 at radii smaller than $\sim25$ kpc, 
but is more extended at radii of $\sim25-100$ kpc. Further at radii 
of $\sim100-200$ kpc, there are no clear differences between N22 
and our result beyond the error bars. The N22 halos seem to be more 
extended than this study at radii larger than 200 kpc, although 
the last bin size of N22 is large ($160-320$ kpc). By this comparison, 
the most notable difference is at radii of $\sim25-100$ kpc.

The reason causing this difference is not clear. The difference is not likely caused by the contamination in our LAEs, 
because the contamination rate of our LAE candidates is low (8\%; see Section \ref{sec:spec}) and we use a median stacking method, the contribution from the contamination is expected to be small ($\lesssim8\%$). Moreover, the contamination is typically foreground point sources that make the surface brightness profile more compact, which cannot explain our more extended profile compared to N22.
There are two possibilities that cause the different results.
The first possibility is the redshift difference. N22 stack LAEs with a larger redshift range ($z=1.9-3.5$) than our LAEs ($z=2.2-2.3$) to increase their sample size, and a possible evolution of LAHs from $z=3$ to 2 may cause the difference. Indeed, it is shown that LAHs at $z=2$ are 0.4 dex fainter than those at $z=3$ at radii smaller than $\sim100$ kpc (\citealt{cai2019}). However, this evolution from $z=3$ to 2 does not greatly change the profile shape (slope), and the redshift evolution alone may not explain the difference between N22 and our results.
The second possibility is the field difference. Our fields are either selected based on a high HI density along background QSO sightlines, or selected to contain multiple QSOs (see \citealt{cai2016}, \citealt{liang2021}, and Cai et al. in prep. for details). As a result, our special field selection may cause the different LAHs.
Note that there are several other stacking studies at $z=2-3$ such as \citet{wisotzki2018} and \citet{kikuchihara2022}. Because results from these studies are similar to M14 and N22, and their surface brightness limits are not as deep as N22, we only compare our results to M14 and N22.

\subsection{Connection between Central Galaxies and LAHs}\label{sec:subsample}
To investigate how central galaxies affect the surrounding LAHs, we divide our LAEs into subsamples based on Ly$\alpha$ luminosity ($L_{\rm Ly\alpha}$), rest-frame Ly$\alpha$ equivalent width (EW$_0$), and UV magnitude (M$_{\rm uv}$). 
If properties ($L_{\rm Ly\alpha}$, EW$_0$, and M$_{\rm uv}$) of a LAE is smaller than the median values, we assign this LAE to the \textit{low} sample. 
On the opposite, we assign a LAE to the \textit{high} sample if its properties 
are larger than the median values. Each subsample thus contains one half of the \textit{all} sample. The median properties of the subsamples are summarized in Table \ref{tab:subsample}.

\begin{deluxetable}{cccc}
\tablenum{2}
\tablecaption{Properties of subsamples\label{tab:subsample}}
\tablewidth{0pt}
\tablehead{
\colhead{Sample} & \colhead{$\log(L_{\rm Ly\alpha})$} & \colhead{EW$_0$} & \colhead{M$_{\rm uv}$}  \\
\colhead{name} & \colhead{erg s$^{-1}$} & \colhead{\AA} & \colhead{mag} 
}
\decimalcolnumbers
\startdata
{\it L\_low} & 42.24 & 59.9 & -18.63 \\
{\it L\_high} & 42.67 & 64.5 & -19.73 \\
{\it EW\_low} & 42.40 & 34.1 & -19.77 \\
{\it EW\_high} & 42.45 & 98.1 & -18.51 \\
{\it Muv\_low} & 42.69 & 38.1 & -20.15 \\
{\it Muv\_high} & 42.30 & 90.5 & -18.53 \\
{\it all} & 42.41 & 52.2 &-19.40 \\
\enddata
\tablecomments{Column 1: sample name; Column 2: median Ly$\alpha$ luminosity; Column 3: median rest-frame Ly$\alpha$ equivalent width; Column 4: median UV magnitude.}
\end{deluxetable}

\begin{figure*}[htb]
\plotone{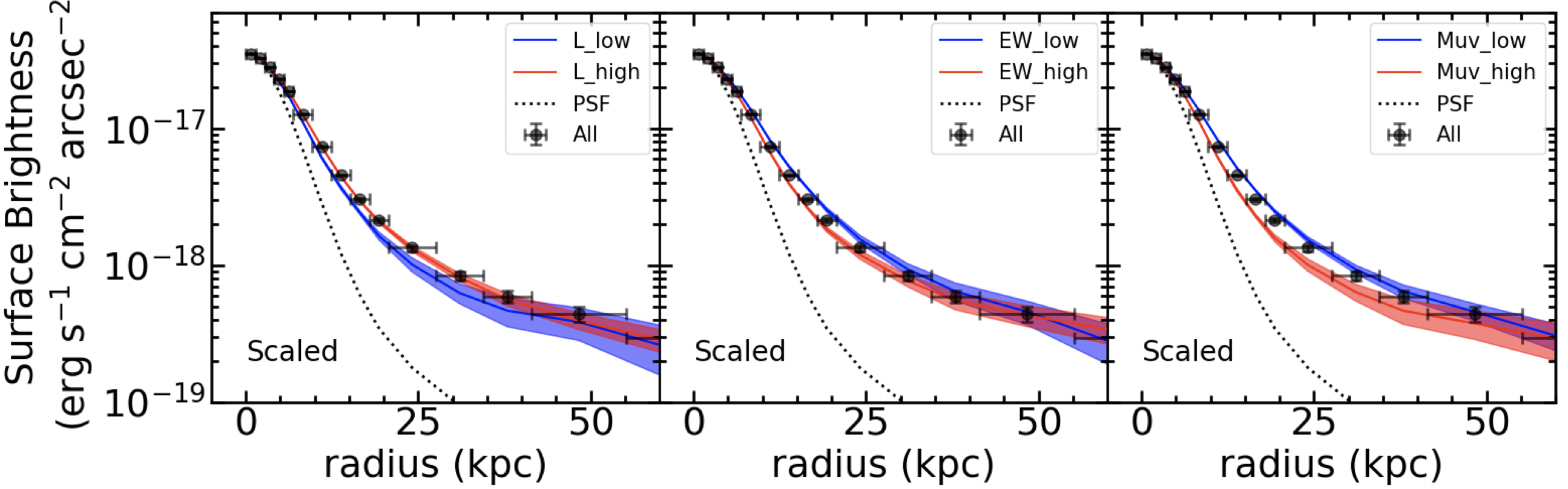}
\caption{Ly$\alpha$ surface brightness profiles of our subsamples (blue for \textit{low} and red for \textit{high}) divided based on $L_{\rm Ly\alpha}$ (left), EW$_0$ (middle), and M$_{\rm uv}$ (right). All of the profiles are scaled at the radius of $\sim0$ kpc for comparison.}
\label{fig:subsample}
\end{figure*}

We stack the subsamples separately using the same method in Section \ref{sec:stacking}. Figure \ref{fig:subsample} shows the stacking results of our subsamples. 
The subsamples show clear differences at the radii of $\sim10-40$ kpc. 
We find that higher $L_{\rm Ly\alpha}$, lower EW$_0$, and brighter M$_{\rm uv}$ cause more extended LAHs. 
Because the three properties are correlated and more massive LAEs generally have 
higher $L_{\rm Ly\alpha}$, lower EW$_0$, and brighter M$_{\rm uv}$ (see {\it Muv\_low} and {\it Muv\_high} subsamples in Table \ref{tab:subsample}), our results suggest that more massive LAEs generally have more extended halos. Similar results have also been shown in \citet{zhang2020}.

\begin{figure}[htb]
\plotone{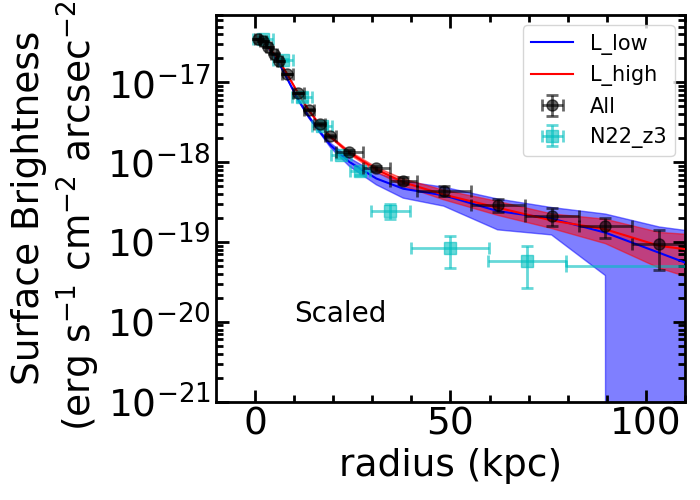}
\caption{Ly$\alpha$ surface brightness profiles of our subsamples (blue for \textit{low} and red for \textit{high}) and \textit{all} sample (black) compared with N22 (cyan). All of the profiles are scaled at the radius of $\sim0$ kpc for comparison.}
\label{fig:subsample2}
\end{figure}

Because LAEs with higher $L_{\rm Ly\alpha}$ generally have more extended halos, 
we investigate whether the halo difference between N22 and our result is caused by 
luminosity differences. 
The median Ly$\alpha$ luminosities of \textit{L\_low}, \textit{L\_high}, and N22 are 42.2, 42.7, and 42.8 erg s$^{-1}$, respectively. Because a higher $L_{\rm Ly\alpha}$ corresponds to a more extended profile, the profile of N22 is expected to be more extended than those of \textit{L\_low} and \textit{L\_high}. However, we find that the profile of N22 is clearly more compact at the radii of $\sim25-100$ kpc, as shown in Figure \ref{fig:subsample2}. This comparison suggests that our more extended Ly$\alpha$ halo compared to N22 cannot be explained by the luminosity difference, but instead by other reasons such as the two possibilities we discussed in Section \ref{sec:stacking}. 

Previous studies have also investigated the connection between central galaxies 
and LAHs as we briefly introduced in Section \ref{sec:intro}. \citet{momose2016} also 
find that galaxies with lower EW$_0$ and brighter M$_{\rm uv}$ have more extended halos. 
However, galaxies with higher $L_{\rm Ly\alpha}$ show less extended halos in \citet{momose2016}, 
which is opposite to this study. \citet{xue2017} find that galaxies with higher $L_{\rm Ly\alpha}$ and brighter M$_{\rm uv}$ have more extended halos, but the halo correlation with EW$_0$ is not clear. It should be noted that the halos in \citet{momose2016} and \citet{xue2017} are only identified at radii smaller than $80$ kpc and have larger uncertainties than this study. Similarly, N22 also show that halo becomes more extended for a higher $L_{\rm Ly\alpha}$, but the relations with EW$_0$ and M$_{\rm uv}$ are not investigated. In summary, the relation between central galaxy properties ($L_{\rm Ly\alpha}$, EW$_0$, and M$_{\rm uv}$) and LAHs in this study is consistent with most previous studies. 

\section{Summary} \label{sec:summary}
In this study, we identify the Ly$\alpha$ halo extended to $\sim200$ kpc by stacking $\sim 3300$ Ly$\alpha$ emitters at $z=2.2-2.3$. Our results are summarized below.

\begin{enumerate}
    \item We carry out imaging observations and data reduction with Subaru/HSC. The total survey area is $\sim12$ deg$^2$ and imaging depths are $25.5-27.0$ mag. Using the imaging data, we select 1240 and 2101 LAE candidates at $z=2.2$ and 2.3, respectively. 
    
    \item We carry out spectroscopic observations of our LAE candidates and data reduction with Magellan/IMACS to estimate the contamination rate of our LAE candidates. We find that the contamination rate of our sample is low (8\%).

    \item We stack our LAE candidates with a median stacking method to identify the Ly$\alpha$ halo at $z=2$. We find that the Ly$\alpha$ halo is extended to $\sim200$ kpc at a surface brightness level of $10^{-20}$ erg s$^{-1}$ cm$^{-2}$ arcsec$^{-2}$. 

    \item Comparing to previous studies, our Ly$\alpha$ halo is consistent with M14 after scaling, but is clearly more extended than N22 at radii of $\sim25-100$ kpc. The halo difference is not likely caused by the contamination in our sample, but by redshift and/or field differences instead.

    \item We divide our LAEs into subsamples based on Ly$\alpha$ luminosity ($L_{\rm Ly\alpha}$), rest-frame Ly$\alpha$ equivalent width (EW$_0$), and UV magnitude (M$_{\rm uv}$). We stack the subsamples and find clear differences between the subsamples at radii of $\sim10-40$ kpc. Our result shows that higher $L_{\rm Ly\alpha}$, lower EW$_0$, and brighter M$_{\rm uv}$ cause more extended halos, which suggests that more massive LAEs generally have more extended Ly$\alpha$ halos.

\end{enumerate}

The Hyper Suprime-Cam (HSC) collaboration includes the astronomical communities of Japan and Taiwan, and Princeton University. The HSC instrumentation and software were developed by the National Astronomical Observatory of Japan (NAOJ), the Kavli Institute for the Physics and Mathematics of the Universe (Kavli IPMU), the University of Tokyo, the High Energy Accelerator Research Organization (KEK), the Academia Sinica Institute for Astronomy and Astrophysics in Taiwan (ASIAA), and Princeton University. Funding was contributed by the FIRST program from Japanese Cabinet Office, the Ministry of Education, Culture, Sports, Science and Technology (MEXT), the Japan Society for the Promotion of Science (JSPS), Japan Science and Technology Agency (JST), the Toray Science Foundation, NAOJ, Kavli IPMU, KEK, ASIAA, and Princeton University. 

This paper makes use of software developed for the Large Synoptic Survey Telescope. We thank the LSST Project for making their code available as free software at  http://dm.lsst.org

The Pan-STARRS1 Surveys (PS1) have been made possible through contributions of the Institute for Astronomy, the University of Hawaii, the Pan-STARRS Project Office, the Max-Planck Society and its participating institutes, the Max Planck Institute for Astronomy, Heidelberg and the Max Planck Institute for Extraterrestrial Physics, Garching, The Johns Hopkins University, Durham University, the University of Edinburgh, Queen’s University Belfast, the Harvard-Smithsonian Center for Astrophysics, the Las Cumbres Observatory Global Telescope Network Incorporated, the National Central University of Taiwan, the Space Telescope Science Institute, the National Aeronautics and Space Administration under Grant No. NNX08AR22G issued through the Planetary Science Division of the NASA Science Mission Directorate, the National Science Foundation under Grant No. AST-1238877, the University of Maryland, and Eotvos Lorand University (ELTE) and the Los Alamos National Laboratory.

Based in part on data collected at the Subaru Telescope and retrieved from the HSC data archive system, which is operated by Subaru Telescope and Astronomy Data Center at National Astronomical Observatory of Japan.

The authors wish to recognize and acknowledge the very significant cultural role and reverence that the summit of Maunakea has always had within the indigenous Hawaiian community.  We are most fortunate to have the opportunity to conduct observations from this mountain.

This paper includes data gathered with the 6.5 meter Magellan Telescopes located at Las Campanas Observatory, Chile.

\appendix
\section{Surface Brightness Distribution of Sky}
\begin{figure*}[htb]
\plotone{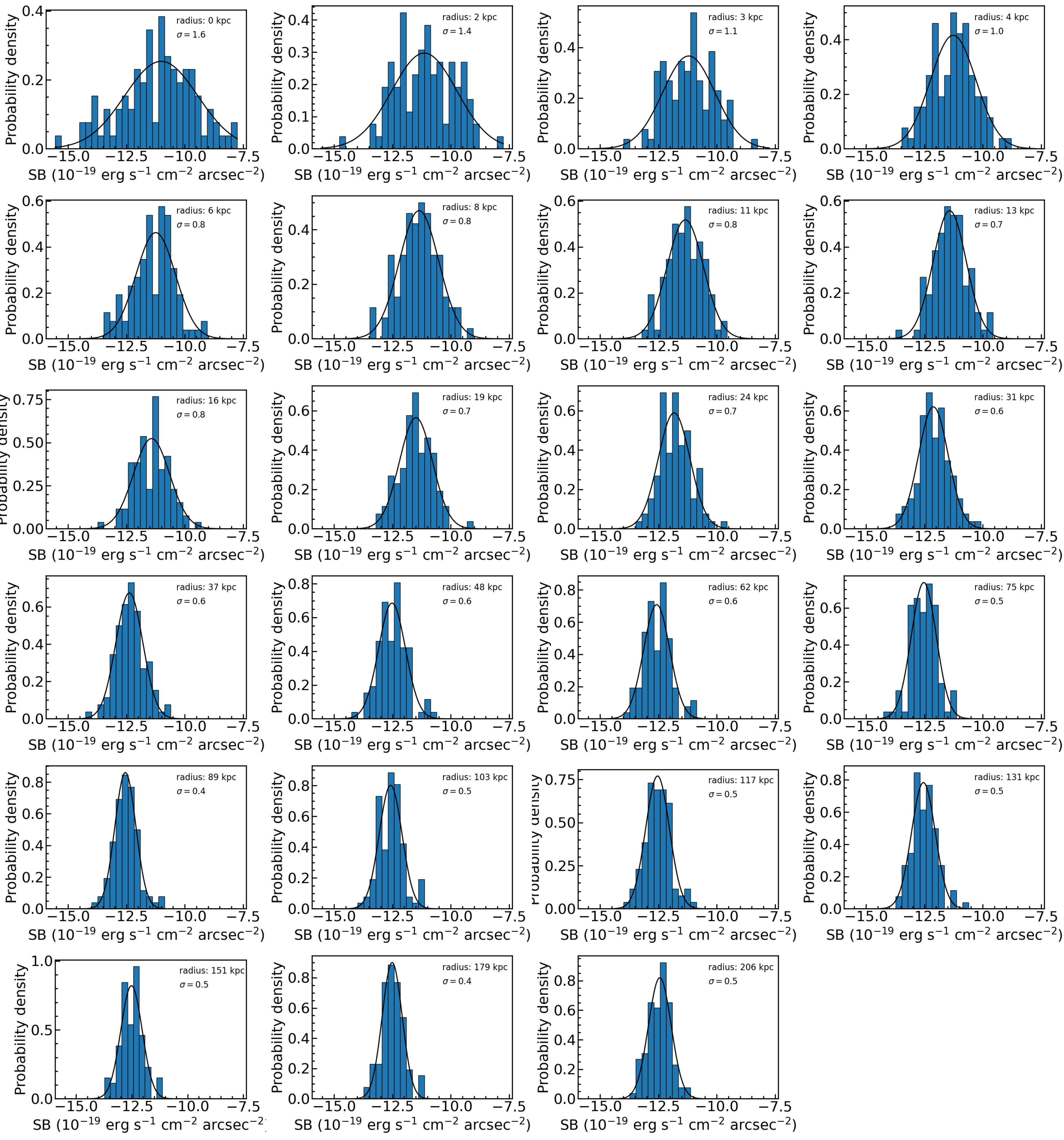}
\caption{Histograms of the surface brightness distribution of sky in each radius bin. The radius of each bin is indicated in each panel. The black solid line shows the best-fit Gaussian function. The $\sigma$ of the best-fit Gaussian function is shown in each panel in a unit of $10^{-19}$ erg s$^{-1}$ cm$^{-2}$ arcsec$^{-2}$. The values of $\sigma$ are equal to the vertical error bars of black points in Figures \ref{fig:stacking_1d}, \ref{fig:halo_compare}, \ref{fig:subsample}, and \ref{fig:subsample2}.}
\label{fig:sky_hist}
\end{figure*}


\begin{thebibliography}{}
\bibitem[Aihara et al.(2018)]{aihara2018} Aihara, H., Arimoto, N., Armstrong, R., et al.\ 2018, \pasj, 70, S4

\bibitem[Arrigoni Battaia et al.(2019)]{battaia2019} Arrigoni Battaia, F., Hennawi, J.~F., Prochaska, J.~X., et al.\ 2019, \mnras, 482, 3162. doi:10.1093/mnras/sty2827

\bibitem[Bertin \& Arnouts(1996)]{bertin1996} Bertin, E. \& Arnouts, S.\ 1996, \aaps, 117, 393. doi:10.1051/aas:1996164

\bibitem[Bosch et al.(2018)]{bosch2018} Bosch, J., Armstrong, R., Bickerton, S., et al.\ 2018, \pasj, 70, S5 

\bibitem[Cai et al.(2016)]{cai2016} Cai, Z., Fan, X., Peirani, S., et al.\ 2016, \apj, 833, 135. doi:10.3847/1538-4357/833/2/135

\bibitem[Cai et al.(2019)]{cai2019} Cai, Z., Cantalupo, S., Prochaska, J.~X., et al.\ 2019, \apjs, 245, 23. doi:10.3847/1538-4365/ab4796

\bibitem[Feldmeier et al.(2013)]{feldmeier2013} Feldmeier, J.~J., Hagen, A., Ciardullo, R., et al.\ 2013, \apj, 776, 75 

\bibitem[Furusawa et al.(2018)]{furusawa2018} Furusawa, H., Koike, M., Takata, T., et al.\ 2018, \pasj, 70, S3

\bibitem[Hayashino et al.(2004)]{hayashino2004} Hayashino, T., Matsuda, Y., Tamura, H., et al.\ 2004, \aj, 128, 2073

\bibitem[Kakuma et al.(2021)]{kakuma2021} Kakuma, R., Ouchi, M., Harikane, Y., et al.\ 2021, \apj, 916, 22. doi:10.3847/1538-4357/ac0725

\bibitem[Kawanomoto et al.(2018)]{kawanomoto2018} Kawanomoto, S., Uraguchi, F., Komiyama, Y., et al.\ 2018, \pasj, 70, 66 

\bibitem[Kikuchihara et al.(2022)]{kikuchihara2022} Kikuchihara, S., Harikane, Y., Ouchi, M., et al.\ 2022, \apj, 931, 97. doi:10.3847/1538-4357/ac69de

\bibitem[Komiyama et al.(2018)]{komiyama2018} Komiyama, Y., Obuchi, Y., Nakaya, H., et al.\ 2018, \pasj, 70, S2

\bibitem[Konno et al.(2016)]{konno2016} Konno, A., Ouchi, M., Nakajima, K., et al.\ 2016, \apj, 823, 20. doi:10.3847/0004-637X/823/1/20

\bibitem[Leclercq et al.(2017)]{leclercq2017} Leclercq, F., Bacon, R., Wisotzki, L., et al.\ 2017, \aap, 608, A8

\bibitem[Liang et al.(2021)]{liang2021} Liang, Y., Kashikawa, N., Cai, Z., et al.\ 2021, \apj, 907, 3. doi:10.3847/1538-4357/abcd93

\bibitem[Lujan Niemeyer et al.(2022)]{niemeyer2022} Lujan Niemeyer, M., Komatsu, E., Byrohl, C., et al.\ 2022, \apj, 929, 90. doi:10.3847/1538-4357/ac5cb8

\bibitem[Magnier et al.(2013)]{magnier2013} Magnier, E.~A., Schlafly, E., Finkbeiner, D., et al.\ 2013, The Astrophysical Journal Supplement Series, 205, 20

\bibitem[Matsuda et al.(2012)]{matsuda2012} Matsuda, Y., Yamada, T., Hayashino, T., et al.\ 2012, \mnras, 425, 878 

\bibitem[Miyazaki et al.(2018)]{miyazaki2018} Miyazaki, S., Komiyama, Y., Kawanomoto, S., et al.\ 2018, \pasj, 70, S1 

\bibitem[Momose et al.(2014)]{momose2014} Momose, R., Ouchi, M., Nakajima, K., et al.\ 2014, \mnras, 442, 110 

\bibitem[Momose et al.(2016)]{momose2016} Momose, R., Ouchi, M., Nakajima, K., et al.\ 2016, \mnras, 457, 2318 

\bibitem[Oke \& Gunn(1983)]{oke1983} Oke, J.~B., \& Gunn, J.~E.\ 1983, \apj, 266, 713 

\bibitem[Rauch et al.(2008)]{rauch2008} Rauch, M., Haehnelt, M., Bunker, A., et al.\ 2008, \apj, 681, 856

\bibitem[Schlafly et al.(2012)]{schlafly2012} Schlafly, E.~F., Finkbeiner, D.~P., Juri{\'c}, M., et al.\ 2012, \apj, 756, 158

\bibitem[Steidel et al.(2011)]{steidel2011} Steidel, C.~C., Bogosavljevi{\'c}, M., Shapley, A.~E., et al.\ 2011, \apj, 736, 160 

\bibitem[Tonry et al.(2012)]{tonry2012} Tonry, J.~L., Stubbs, C.~W., Lykke, K.~R., et al.\ 2012, \apj, 750, 99

\bibitem[Wisotzki et al.(2016)]{wisotzki2016} Wisotzki, L., Bacon, R., Blaizot, J., et al.\ 2016, \aap, 587, A98 

\bibitem[Wisotzki et al.(2018)]{wisotzki2018} Wisotzki, L., Bacon, R., Brinchmann, J., et al.\ 2018, \nat, 562, 229 

\bibitem[Xue et al.(2017)]{xue2017} Xue, R., Lee, K.-S., Dey, A., et al.\ 2017, \apj, 837, 172. doi:10.3847/1538-4357/837/2/172

\bibitem[Zhang et al.(2020)]{zhang2020} Zhang, H., Ouchi, M., Itoh, R., et al.\ 2020, \apj, 891, 177. doi:10.3847/1538-4357/ab7917


\end{thebibliography}
\end{document}